\documentclass[12pt,preprint]{aastex}

\newcommand{\etal}{et~al.}
\newcommand{\fig}{Fig.~}
\newcommand{\mic}{\,$\mu$m}

\newcommand{\av}{A_V}
\newcommand{\cod}{CO$_2$}
\newcommand{\water}{H$_2$O}
\newcommand{\methanol}{CH$_3$OH}

\slugcomment{Version date: \today}

\shorttitle{CO$_2$ paper}
\shortauthors{Whittet et al.}

\begin{document}

\title{THE NATURE OF CARBON DIOXIDE BEARING \\ ICES IN QUIESCENT MOLECULAR CLOUDS\\~\\}

\author{D.\ C.\ B. Whittet\altaffilmark{1}, A.\ M. Cook\altaffilmark{1}, J.\ E. Chiar\altaffilmark{2},\\ 
Y.\ J. Pendleton\altaffilmark{3}, S.\ S. Shenoy\altaffilmark{4}, and P.\ A. Gerakines\altaffilmark{5} \\~\\}

\altaffiltext{1}{Department of Physics, Applied Physics \& Astronomy,
	             Rensselaer Polytechnic Institute, 110 Eighth Street, Troy, NY 12180.}
\altaffiltext{2}{SETI Institute, 515 N.\ Whisman Road, Mountain View, CA 94043.}
\altaffiltext{3}{NASA Ames Research Center, Mail Stop 245-1, Moffett Field, CA 94035.}
\altaffiltext{4}{Spitzer Science Center, Mail Code 220-6, California Institute of Technology, Pasadena, CA 91125.}
\altaffiltext{5}{Department of Physics, University of Alabama at Birmingham,
	             310 Campbell Hall, 1300 University Blvd., Birmingham, AL 35294.}

\begin{abstract}
The properties of the ices that form in dense molecular clouds represent an important set of initial conditions in the evolution of interstellar and preplanetary matter in regions of active star formation. Of the various spectral features available for study, the bending mode of solid \cod\ near 15\mic\ has proven to be a particularly sensitive probe of physical conditions, especially temperature. We present new observations of this absorption feature in the spectrum of \hbox{Q21--1}, a background field star located behind a dark filament in the Cocoon Nebula (IC~5146). We show the profile of the feature be consistent with a two-component (polar + nonpolar) model for the ices, based on spectra of laboratory analogs with temperatures in the range 10--20~K. The polar component accounts for $\sim 85$\% of the \cod\ in the line of sight. We compare for the first time 15\mic\ profiles in three widely separated dark clouds (Taurus, Serpens and IC~5146), and show that they are indistinguishable to within observational scatter. Systematic differences in the observed \cod/\water\ ratio in the three clouds have little or no effect on the 15\mic\ profile. The abundance of elemental oxygen in the ices appears to be a unifying factor, displaying consistent behavior in the three clouds. We conclude that the ice formation process is robust and uniformly efficient, notwithstanding compositional variations arising from differences in how the O is distributed between the primary species (\water, \cod\ and CO) in the ices.  

\end{abstract}

\keywords{Dust, extinction --- Infrared: ISM: lines and bands \\ --- ISM: molecules}

%\clearpage 

\section{Introduction}
Dust grains acquire icy mantles in the cold, dense molecular phase of the  (ISM), the composition of which is determined both by gas-phase chemistry and by chemical reactions occurring on the surfaces of the grains (e.g., Tielens \& Hagen 1982). Interstellar ices incorporated into the envelopes and disks of young stellar objects (YSOs) may subsequently be heated, chemically or structurally modified or destroyed by energetic processes such as exposure to radiation, shocks and winds. The study of ices in the environments of YSOs is currently a highly active field, driven by the rich legacy of infrared spectra available from the Infrared Space Observatory (ISO) and the Spitzer Space Telescope (e.g., de~Graauw \etal\ 1996; d'Hendecourt \etal\ 1996; Gerakines \etal\ 1999; Nummelin \etal\ 2001; Gibb \etal\ 2004; Boogert \etal\ 2004, 2008; Watson \etal\ 2004; Dartois \etal\ 2005; Pontoppidan \etal\ 2005, 2008; Zasowski \etal\ 2008). Of the various infrared absorption features identified with vibrational modes of \water, CO, \cod\ and other molecular species in the ices, the O=C=O bending mode of \cod\ near 15\mic\ has proven to be particularly useful as a tool sensitive to physical conditions, especially temperature. Ices formed and maintained at low temperature are amorphous in structure, but studies of laboratory analogs have demonstrated the appearance of subfeatures in the 15\mic\ profile upon heating that are diagnostic of segregation and crystallization of constituents (Ehrenfreund \etal\ 1997, 1999; Dartois \etal\ 1999; van Broekhuizen \etal\ 2006; White \etal\ 2009). These subfeatures have, indeed, been observed in the infrared spectra of a variety of high- and low-mass YSOs (see Pontoppidan \etal\ 2008 for numerous examples). 

The properties of pristine ices in cold, quiescent regions of dense molecular clouds provide an important benchmark for studies of the evolution of interstellar matter and the origins of stars and planetary systems. To study ices in such regions, it is necessary to observe background field stars viewed through the dense material, and this is much more difficult compared with YSOs because of their relatively weak intrinsic flux at mid-infrared wavelengths. First detection of \cod-ice in a field star resulted from ISO observations of its near infrared (4.2\mic) stretching-mode feature in the spectrum of the Taurus K-type giant Elias~16 (Whittet \etal\ 1998). Subsequent observations with Spitzer confirmed the presence of the bending-mode feature in this line of sight (Bergin \etal\ 2005; Knez \etal\ 2005), exhibiting a profile that is, as expected, notably lacking in structure associated with heating and crystallization of the ices. The feature is consistent with models that utilize data for laboratory analogs at low temperature \hbox{($T\sim 10$--20~K)}, combining `polar' (\water-rich) and `non-polar' (\water-poor) components (e.g.\ Gerakines \etal\ 1999). Detections of \cod\ in the spectra of a further 8~field stars behind the Taurus dark-cloud complex have since been reported (Whittet \etal\ 2007 and references therein). Consistency was found both in the \cod\ abundance ($\sim 18\%$ relative to \water) and in the 15\mic\ profile shape: each line of sight displays a profile essentially identical to that of Elias~16 to within observational error. This result implies that the ices in Taurus have been maintained at low temperature since their formation, and furthermore that the formation process is fairly robust and uniformly efficient at several locations in the cloud.

Taurus has (by default) become the prototype for study of ices in low-mass star-formation regions because of the relative ease with which they are detected in this dark-cloud complex\footnote{The Taurus complex, which includes TMC--1, L1495, B18 and several other dense cores, is relatively nearby, with little foreground extinction. Ice features are detected at relatively low extinction compared with other clouds studied to date and are widely distributed across the region (see Shenoy \etal\ 2008 for further discussion and references).}. A key question is whether it is a {\it reliable\/} prototype. Substantial cloud-to-cloud variations are known to exist, for example, in ambient radiation field (Liseau \etal\ 1999) and grain temperature (White \etal\ 1995), and also in gas phase chemistry between clouds of similar temperature, density and radiative environment (Dickens \etal\ 2000). It is important to determine whether such variations have an appreciable impact on the composition and physical state of the ices in order to test the validity of Taurus as a prototype. However, background field stars bright enough at 15\mic\ to yield spectra of sufficient quality and with enough extinction to show deep interstellar features are rare. To date the profile of the sensitive \cod\ bending mode has been reported in only one confirmed background field star outside of the Taurus region: CK2 in Serpens (Knez \etal\ 2005). The Serpens dark cloud differs from that in Taurus in both structure and content: it is more centrally condensed, with higher average gas densities, and displays greater intermediate-mass star formation activity leading to warmer ambient gas temperatures ($\sim 25$K compared with $\sim 15$K; White \etal\ 1995; McMullin \etal\ 2000; Kaas \etal\ 2004). There is evidence also to suggest differences in ice composition between the two clouds, with (e.g.) an excess of CO and/or \cod\ relative to \water\ in Serpens compared with Taurus (Eiroa \& Hodapp 1989; Chiar \etal\ 1994; Whittet \etal\ 2007).

In this paper, we report observations and profile modeling of the \cod\ bending-mode ice feature in a field star background to a dark, filamentary cloud associated with the Cocoon Nebula (IC~5146) in Cygnus (the ``Northern Streamer'' mapped by Lada \etal\ 1994). We compare results for the three regions now available --- Taurus, Serpens and IC~5146 --- and show that they do, indeed, support the existence of a common 15\mic\ ice profile for unprocessed ices in the cold dark-cloud environment. Our results also allow new constraints to be placed on the ice formation process.

\section{Spectra}
All \cod\ spectra used in this work were obtained with the Infrared Spectrograph (IRS) of the Spitzer Space Telescope, described by Houck \etal\ (2004). The IRS operated in short-wavelength, high-resolution staring mode, covering the spectral range 10.0--19.5\mic\ at a resolving power $\lambda/\Delta\lambda \approx 600$. The Astronomical Observation Request (AOR) keys and other relevant information are given in Table~1 for our three program stars. The spectrum of Q21--1 (IC~5146), reported here for the first time, was obtained on 2004 November~12 as part of the Spitzer General Observer program \#3320 (Principal Investigator: Y.~Pendleton); three ramp cycles were carried out to yield a total on-source integration time of 7975\,s, and a spectrum of the sky background in the region was also obtained to provide background subtraction. The raw spectra were processed by the standard Spitzer pipeline to produce basic calibrated data (BCD). Version 15.3.0 of the pipeline was used to reduce the data for Q21--1 and also for CK2 (a reduction of the CK2 spectrum using pipeline version 11.0.2 was previously published by Knez \etal\ 2005). For Elias~16, we use the spectrum reported by Bergin \etal\ (2005), processed with version~13.0 of the pipeline\footnote{Tests showed no improvement in data quality using pipeline version~15.3 to process the spectrum for Elias~16 compared with the published spectrum.}. Further analysis of the BCD to produce final calibrated spectra of Q21--1 and CK2 was carried out using the SMART program (Higdon \etal\ 2004) and standard procedures to remove bad pixels and other artifacts, as described in \S2 of Whittet \etal\ (2007). 

The final calibrated flux spectrum of Q21--1 is shown in \fig 1a. A local continuum was determined by fitting a second-order polynomial to adjacent wavelength regions on each side of the \cod\ feature (13.0--14.9\mic\ and 16.1--17.6\mic). The adopted continuum, shown as a dashed curve in \fig 1a, was used to calculate the optical depth spectrum plotted in \fig 1b. The same procedure was used to produce an optical depth spectrum for CK2 (see below); our result for CK2 is broadly consistent with that reported by Knez \etal\ (2005), allowing for differences in the adopted continuum.

\section{Results}
The final optical depth profiles of the \cod\ features observed in CK2 and Q21--1 are shown in \fig 2, together with fits based on laboratory analogs (Table~2). \fig 3 (lower frame) compares the profiles of CK2 and Q21--1 with that of Elias~16, superposed with normalization to unity at peak optical depth. It is evident from inspection that the observed profiles in the three lines of sight are closely similar, indeed virtually identical to within observational error. This conclusion is supported by the random nature of the scatter in the residual optical depths for CK2 and Q21--1 relative to Elias~16, shown in the upper frame of \fig 3.

\subsection{Comparison with laboratory analogs}
The fitting procedure used to model the \cod\ profile is the same as described in our previous work (e.g.\ Gerakines \etal\ 1999; Bergin \etal\ 2005), but with access to a much larger suite of analog spectra now available. The procedure assumes that the profile can be modeled by a combination of two ice components\footnote{The choice of two as the number of components is justified as the smallest capable of giving an acceptable fit (no single mixture can reproduce the profile), but it should be noted that as many as five components have sometimes been used to fit \cod\ bending-mode profiles in YSOs (Pontoppidan \etal\ 2008).}: one contains \cod\ in a matrix dominated by \water\ (the polar component), the other contains CO and \cod, but little or no \water\ (the nonpolar component). Some analogs also contain methanol (\methanol), which can affect the shape of the \cod\ bending mode (Ehrenfreund \etal\ 1999). A total of 1381~laboratory spectra is available from the databases of the Astrophysics Laboratories of Leiden University (Ehrenfreund \etal\ 1997, 1999) and the University of Alabama at Birmingham (White \etal\ 2009), spanning a wide range of temperature \hbox{(5--150~K)} as well as composition. 

We fit each spectrum independently to further test the hypothesis that the profiles are effectively the same. Best fits resulting from our least-squares fitting procedure are summarized in Table~2 and plotted in \fig 2 for CK2 and Q21--1 (our fit to Elias~16 is similar to that shown in \fig 2 of Bergin \etal\ 2005). As expected, the combinations of ices selected are qualitatively and quantitatively similar, with temperatures in the range 10--20~K; indeed the same polar ice mixture was selected in each case, and the nonpolar mixtures differ only in CO:\cod\ ratio. In fact, the fits are not very sensitive to \cod\ concentration in either component\,--- different mixtures give almost equally good fits (see also \S4.1 below) --- but they are robust in terms of three essential requirements: (i)~both polar and nonpolar components must be present, (ii)~both components must be at low temperature ($<25$~K), and (iii)~the profile is dominated by the polar component, which contributes $\sim 80$--90\% of the total absorption in each line of sight. None of the \methanol-bearing mixtures were selected by the fitting process, consistent with available limits on its solid-state abundance in dark clouds (Chiar \etal\ 1996).

\subsection{Column densities}
The integrated optical depth, $\int\tau(\nu)d\nu$, of the observed absorption feature in each program star was estimated, and results are listed in Table~1. This quantity is directly related to column density:
\begin{equation}
N = {\int\tau(\nu)d\nu\over A}
\end{equation}
where $A$ is the band strength. Gerakines \etal\ (1995) showed that the band strength of the \cod\ bending mode is only weakly dependent on the ice mixture in which the \cod\ resides, and we therefore adopt an average value of $A = 1.0 \times 10^{-17}$~cm/molecule, consistent with our previous work (Whittet \etal\ 2007). Column densities for \cod\ are listed in Table~2, together with values for other major constituents of the ices (\water\ and CO) from the literature. All $N$(\water) values are determined from ground-based observations of the 3.0\mic\ O--H stretching mode, which generally gives more reliable results than the blended O--H--O bending mode near 6\mic.

\fig 4 plots CO and \cod\ ice column densities against that of \water. In addition to Elias~16, CK2 and Q21--1, all Taurus field stars with available data are included (Whittet \etal\ 2007 and references therein). The straight lines are linear least-squares fits to Taurus stars only. The fit to $N$(CO) vs.\ $N$(\water) displays a positive intercept on the $N$(\water)-axis, as expected, as \water\ is less volatile and can exist in solid form at lower dust columns compared with CO. It is evident (\fig 4a) that both CK2 in Serpens and Q21--1 in IC~5146 have significantly higher \cod/\water\ abundance ratios compared with members of the Taurus group (which show a rather tight correlation). Systematic differences are less apparent in the plot of $N$(CO) vs.\ $N$(\water), suggestive of a common trend (\fig 4b) albeit with appreciable scatter; the locus of Q21--1, in particular, is closely consistent with the Taurus correlation. 
 
\section{Discussion}
The results described in the previous section demonstrate that the \cod\ ice feature near 15\mic\ displays a common profile shape in three widely separated dense clouds. The available spectra are consistent with a combination of polar (\water-rich) and nonpolar (CO-rich) ices at low temperature (10--20~K), the polar component accounting for $\sim 85$--90\% of the \cod. Any elevation in ice temperature arising from proximity to embedded stars (e.g., Serpens: see \S1) must not be sufficient to affect the profile in the lines of sight studied. Interestingly, this commonality of profile also occurs notwithstanding a clear difference in the composition of the ices, manifested by discontinuity in the $N$(\cod) vs.\ $N$(\water) correlation for IC~5146 and Serpens compared to Taurus (\fig 4a). In this section, we explore possible implications of these results. We hypothesize that the compositional discontinuity between clouds may arise from variations in only one parameter: the \cod/\water\ ratio in the polar ices. 

\subsection{Stability of the profile}
It was noted previously that the bending-mode absorption profiles of ices containing \cod\ are not strongly sensitive to \cod\ concentration at a given temperature. This remains true provided only that the ices maintain their compositional and structural `identity' (e.g., a polar mixture contains enough \water\ to remain `polar'). For illustration, we compare in \fig 5 the profiles for several polar laboratory ice mixtures at low temperature, with \cod\ concentration varying between 14\% and 100\% relative to \water. Slight differences in position and width are seen, but the profile shape is maintained over this rather broad compositional range. It is therefore unsurprising that observed profiles are closely similar in lines of sight with different \cod/\water\ ratios.

\subsection{Oxygen budget}
Polar ices are thought to form on interstellar grains by surface reactions involving species (principally H, O and CO) adsorbed from the gas phase in the ``diffuse-dense" transition region of a molecular cloud where H $\rightarrow$ H$_2$ conversion is incomplete. In brief, the main chemical pathways appear to be hydrogenation of O to OH and \water, and oxidation of CO to \cod, with no more than trace amounts of the CO undergoing hydrogenation to form CHO-bearing species such as \methanol\ (e.g., Tielens \& Whittet 1997)\footnote{Note that the chemistry leading to \cod\ formation is not yet fully understood because of uncertainties in the rates of key surface reactions such as CO + O $\rightarrow$ \cod\ and CO + OH $\rightarrow$ \cod\ + H; see Ruffle \& Herbst (2001) and Pontoppidan \etal\ (2008) for further discussion.}. The \cod/\water\ ratio in the resulting solid is likely to depend on the relative abundances of the adsorbed species, and thus on the network of chemical reactions occurring in the gas phase. Local conditions may lead to variations. A cloud with a relatively high concentration of gas phase CO relative to atomic H in the transition region will tend to produce a more \cod-rich polar ice, whereas a cloud with a lower concentration of gas phase CO will produce less \cod\ and free up more O to form \water: the differences evident in \fig 4a between Taurus and the other clouds might be explained in this way. At higher densities, H $\rightarrow$ H$_2$ conversion is essentially complete and \water\ can no longer form efficiently by surface reactions; nonpolar ices then accumulate on the grains, principally by attachment of gas-phase CO (Pontoppidan 2005), which is abundant in the dense molecular phases of all dark clouds. Formation of \cod\ within the nonpolar component may be driven by cosmic-ray processing of the solid CO (Pontoppidan \etal\ 2008).

It is informative to consider the distribution of elemental oxygen in the ices arising from these processes. Ices, together with silicate dust and gas-phase CO, represent the major observable oxygen reservoirs in the dense ISM (Whittet \etal\ 2007 and references therein). O is the most abundant heavy element and the only one present in all three of the major molecular ice species, \water, CO and \cod. Atomic O attaching to a grain surface will undergo chemical reactions that lead to formation of \water\ and \cod, as outlined above: {\it we hypothesize that the fraction of O atoms accumulating into the ices is robust but that their subsequent distribution between \water\ and \cod\ may vary from cloud to cloud.}

The total number of O atoms in the polar ices along a line of sight may be expressed in terms of column density by the relation
\begin{equation} 
N({\rm O})_{\rm polar} = N({\rm H_2O}) + f_1 N({\rm CO}) + 2 f_2 N({\rm CO_2}),
\end{equation}
where $f_1$ and $f_2$ are the fractions of all solid-phase CO and \cod\ molecules, respectively, in the polar ices. Similarly, for the nonpolar ices, we have
\begin{equation} 
N({\rm O})_{\rm nonpolar} = (1-f_1) N({\rm CO}) + 2(1-f_2) N({\rm CO_2}).
\end{equation}
Previous and current work suggests $f_1\approx 0.17$ in Serpens and $0.12$ in Taurus and IC~5146 (Chiar \etal\ 1994, 1995 and in preparation); $f_2$ is evaluated from the results of the present paper (Table~2, right hand column). We ignore other species that might contain significant elemental oxygen in the ices, of which the most important is probably O$_2$ (see Whittet \etal\ 2007 for further discussion).

\fig 6 plots $N$(O) for polar and nonpolar ices against visual extinction ($\av$). As in \fig 4, all available Taurus field stars are included using data from Whittet \etal\ (2007) in addition to results from the present work (Elias~16 has the highest $\av$ value of the Taurus stars plotted). $\av$ is taken as a measure of total dust column. High degrees of correlation are evident in \fig 6, with notable consistency between the three clouds for both polar and nonpolar ices: in contrast to the situation in \fig 4, a single linear fit to all points is representative of the trend for each component of the ices. This result suggests stability in the net rate of accumulation of O atoms into ices on grain surfaces, regardless of differences in the efficiencies of chemical pathways that form \water\ and \cod.

Some caveats should be noted. First, the extinction toward CK2 is particularly uncertain: we have adopted the value from Knez \etal\ (2005), but higher values have been estimated in earlier literature (see discussion in \S6 of Whittet \etal\ 2007). Second, we are concerned only with the component of $\av$ arising in dense clouds (where grains are ice-mantled); any systematic cloud-to-cloud differences in the extinction contributed by unmantled grains in diffuse phases of the ISM along these lines of sight would lead to systematic error in the form of a horizontal shift in \fig 6. This contribution may be estimated empirically by evaluating the ice `threshold' extinction, i.e., the intercept on the $\av$ axis of fits to correlations of \water-ice optical depth or column density with visual extinction: a value of $\sim 3.3$~mag is well documented for the Taurus cloud (Whittet \etal\ 1988, 2001), and this is, indeed, consistent with the locus of the intercept of the linear fit to the polar component in \fig 6. Available data suggest somewhat higher thresholds for IC~5146 ($\sim 4.0$~mag; Chiar \etal\ in preparation) and Serpens ($\sim 5.5$~mag; Eiroa \& Hodapp 1989). However, in each case, the difference relative to the Taurus value is smaller than the estimated uncertainty in $\av$ for Q21--1 and CK2, respectively. Therefore, we do not consider this potential source of error to be of major significance.

\fig 6 suggests a uniform distribution of elemental oxygen between polar and nonpolar components, the latter accounting for $\sim 20\%$ of the total in the ices at high $\av$. This conclusion is supported by \fig 7, which plots $N$(O) for polar ices against that for nonpolar ices (eqs.~2 and 3). \fig 7 is unaffected by any systematic error introduced by cloud-to-cloud difference in diffuse-ISM extinction, and may be compared with \fig 4b (the largest contributions to O in polar and nonpolar ices being made by \water\ and CO, respectively). The data are consistent with a linear correlation with continuity between clouds. The positive intercept of the fit on the $N$(O)$_{\rm polar}$ axis in \fig 7 is probably real (c.f.\ \fig 4b) and consistent with the lower volatility of polar ices compared with nonpolar ices. 

\subsection{Homogeneity of polar/nonpolar ice phases}
Consistency in the distribution of matter between polar and nonpolar phases of the ice is indicated both by the correlations in Figs.\,6 and 7 and by the results of the 15\mic\ profile fitting (\S3.1). An unexpectedly high level of compositional homogeneity is implied: no major, systematic variations in the relative abundance of polar to nonpolar ice is apparent over a rather wide range of environments and extinctions ($5\la \av \la 35$~mag; \fig 6). Polar ices form at relatively low density, as discussed above, and their abundance is expected to correlate linearly with dust column (i.e., with $\av$) if mantled grains are widely distributed throughout the clouds. In the case of nonpolar ices, however, the abundance of the primary species (CO) is not dependent on formation in situ but rather on accumulation from the gas. The rate of accumulation would thus be expected to increase rapidly with density at high $\av$ unless there is an efficient desorption mechanism. Our results suggest that a balance is maintained between accretion and desorption of CO. This situation is, indeed, entirely consistent with observations of gas-phase CO in molecular clouds in general (substantial levels of gas phase CO are maintained in dense regions), and with laboratory work on the photodesorption properties of CO ice (see \"Oberg \etal\ 2007 for further discussion).

\section{Conclusions}
The primary result of this study is a clear demonstration that the \cod\ ice bending-mode feature near 15\mic\ displays essentially the same profile shape in lines of sight that intercept three widely separated dense clouds. This profile is likely to be a useful benchmark for investigations of \cod-bearing ices in YSOs, aiding separation of contributions made by the circumstellar environment and the parent molecular cloud to each observed spectrum. 

The shape of the observed profile elucidates the nature of the ices in the dense-cloud environment. Comparisons with laboratory analogs confirm that the \cod-bearing ices are maintained at low temperature \hbox{($< 25$~K)} and include both polar (\water-dominated) and nonpolar (CO-dominated) components. The polar component contains $\sim 85$--90\% of all solid-phase \cod, but the concentration of \cod\ relative to \water\ is not tightly constrained. The constancy of the profile in the face of observed differences in the \cod/\water\ abundance ratio is thus easily understood. 

Cloud to cloud variations in \cod/\water\ ratio may result from differences in gas phase chemistry, but the physical process of element depletion from the gas onto dust proceeds in a predictable manner. The elemental oxygen column density in the ices appears to be a unifying parameter, showing evidence of consistent behavior with respect to dust column in our sample. Moreover, the distribution of elemental oxygen between polar and nonpolar ice appears to be almost invariant: the CO-rich nonpolar component accounts for no more than $\sim 20\%$, even in the densest regions sampled, a result which implies the existence of an efficient desorption mechanism for CO in molecular cloud cores. 

It will be important in the future to extend this work to a larger sample of field stars that probe quiescent regions of molecular clouds. Given the constancy of profile observed to date, spectra obtained with the low-resolution mode of the Spitzer IRS may be adequate to extend abundance correlations such as those shown in Figs.\,4 and 6, allowing fainter stars to be included. Opportunities to obtain high-quality spectra of other ice constituents best observed from airborne and/or space platforms will be provided in the next few years by the availability of the Stratospheric Observatory for Infrared Astronomy and the James Webb Space Telescope. 

\acknowledgments
This work is based on observations made with the Spitzer Space Telescope, which is operated by the Jet Propulsion Laboratory, California Institute of Technology, under a contract with NASA. The authors acknowledge NASA support from grant NNX07AK38G and JPL/Caltech Support Agreement no.\,1290823 (D.C.B.W.) and JPL/Caltech subcontracts 1266411 and 1267778 (J.E.C.). We are grateful to an anonymous referee for helpful comments that led to improvements to this paper.

%{\it Facilities:} \facility{Spitzer (IRS)}

\clearpage

\begin{deluxetable}{lllccccc} 
\tabletypesize{\scriptsize} 
\tablecaption{\small Summary of observations and related information for program stars. \label{table1}}
\tablewidth{0pt} 
\tablehead{\colhead{2MASS ID} & \colhead{Other ID} & \colhead{Cloud} & \colhead{Spectral} & \colhead{$A_V$} 
	& \colhead{AOR key} & \colhead{{$\int\tau(\nu)d\nu$}$^{\rm a}$} & \colhead{Notes\,$^{\rm b}$}\\ 
& & & type & (mag) & & (cm$^{-1}$)}
\startdata 
J04393886+2611266 & Elias 16     & Taurus  & K1 III & $24.1 \pm 0.5$ & 03868160 & $5.4 \pm 0.2$ & (1) \\ 
J18300061+0115201 & CK2, EC\,118 & Serpens & K4 III &   $34 \pm 4$   & 11828224 & $9.9 \pm 0.4$ & (2) \\ 
J21472204+4734410 & Q21--1       & IC 5146 & K2 III &   $27 \pm 1$   & 10751488 & $8.7 \pm 0.4$ & (3) \\ 
\enddata 
\tablenotetext{a}{Integration was carried out over the wavelength range 14.9--16.1\mic.}
\tablenotetext{b}{Notes:
(1)~IRS spectrum from Bergin \etal\ 2005; spectral type from Elias 1978; visual extinction ($A_V$) from Shenoy \etal\ 2008. (2)~IRS spectrum previously reported by Knez \etal\ 2005 and rereduced by us (see \S2); spectral type and extinction from Knez \etal\ 2005. (3)~IRS spectrum from the current work (\S2); spectral type and extinction from Chiar \etal, in preparation.} 
\end{deluxetable}

\begin{deluxetable}{lcccllcc} 
\tabletypesize{\scriptsize} 
\tablecaption{\small Column densities and details of fits to the \cod\ bending-mode feature. \label{table2}}
\tablewidth{0pt} 
\tablehead{\colhead{Star} & \multicolumn{3}{c}{Column densities$^a$} & \colhead{Component} & \colhead{Mixture} 
	& \colhead{$T$} & \colhead{Percent}\\
	& \colhead{$N$(H$_2$O)}& \colhead{$N$(CO)} & \colhead{$N$(CO$_2$)} &&& \colhead{(K)} & \colhead{$N$(\cod)}}
\startdata 
Elias 16 & 25.7\,$^{\rm b}$ & 6.5\,$^{\rm c}$ & 5.4 & polar     &  \water:\cod:CO = 100:20:3 & 20 & 89\% \\ 
         &                  &                 &     & nonpolar  &  CO:\cod\       = 100:23   & 10 & 11\% \\
CK2      & 35\,$^{\rm d}$   &16.2\,$^{\rm e}$ & 9.9 & polar     &  \water:\cod:CO = 100:20:3 & 20 & 88\% \\ 
         &                  &                 &     & nonpolar  &  CO:\cod\       = 100:26   & 10 & 12\% \\
Q21--1   & 25.2\,$^{\rm f}$ & 7.1\,$^{\rm f}$ & 8.7 & polar     &  \water:\cod:CO = 100:20:3 & 20 & 85\% \\ 
         &                  &                 &     & nonpolar  &  CO:\cod\       = 100:70   & 10 & 15\% \\
\enddata 
\tablenotetext{a}{All column densities are in units of $10^{17}$~cm$^{-2}$; all $N$(CO$_2$) values are from the current work.} 
\tablenotetext{b}{Average value from 3\mic\ spectra of Whittet \etal\ 1988, Smith \etal\ 1993 and Murakawa \etal\ 2000.}
\tablenotetext{c}{Chiar \etal\ 1995.}
\tablenotetext{d}{Calculated from 3\mic\ spectrum of Eiroa \& Hodapp 1989.}
\tablenotetext{e}{Average of values from Chiar \etal\ 1994 and Pontoppidan \etal\ 2003 (as cited by Knez \etal\ 2005).}
\tablenotetext{f}{From unpublished observations (Chiar et~al., in preparation).}
\end{deluxetable}

\clearpage

% Fig.1
\begin{figure}
\epsscale{0.75}
\plotone{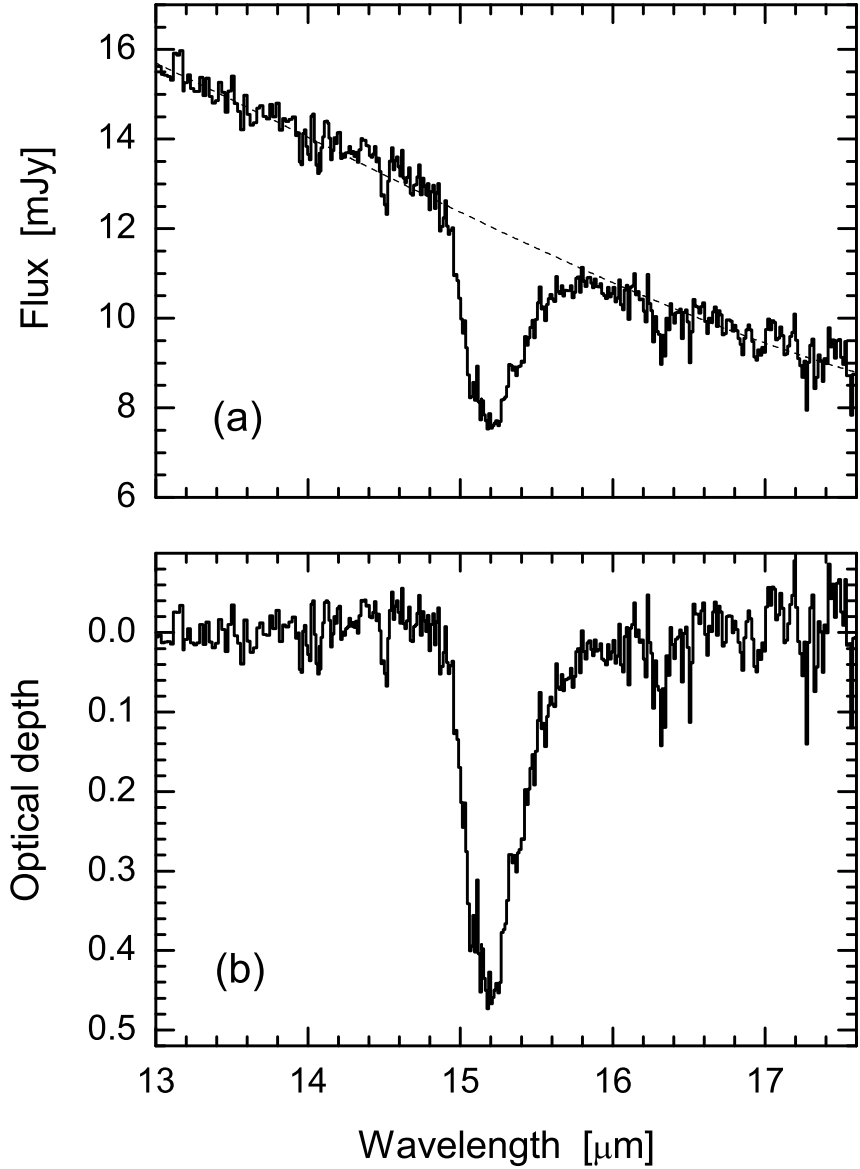}
\caption{(a) Flux spectrum of Q21--1 (IC~5146) in the spectral range 13.0--17.6\mic. The continuum adopted for the \cod\ feature centered near 15.2\mic\ is shown as a dashed curve. (b) Optical depth spectrum over the same spectral range, obtained using the continuum shown in (a).\label{fig1}}
\end{figure}

\clearpage

% Fig.2
\begin{figure}
\epsscale{0.85}
\plotone{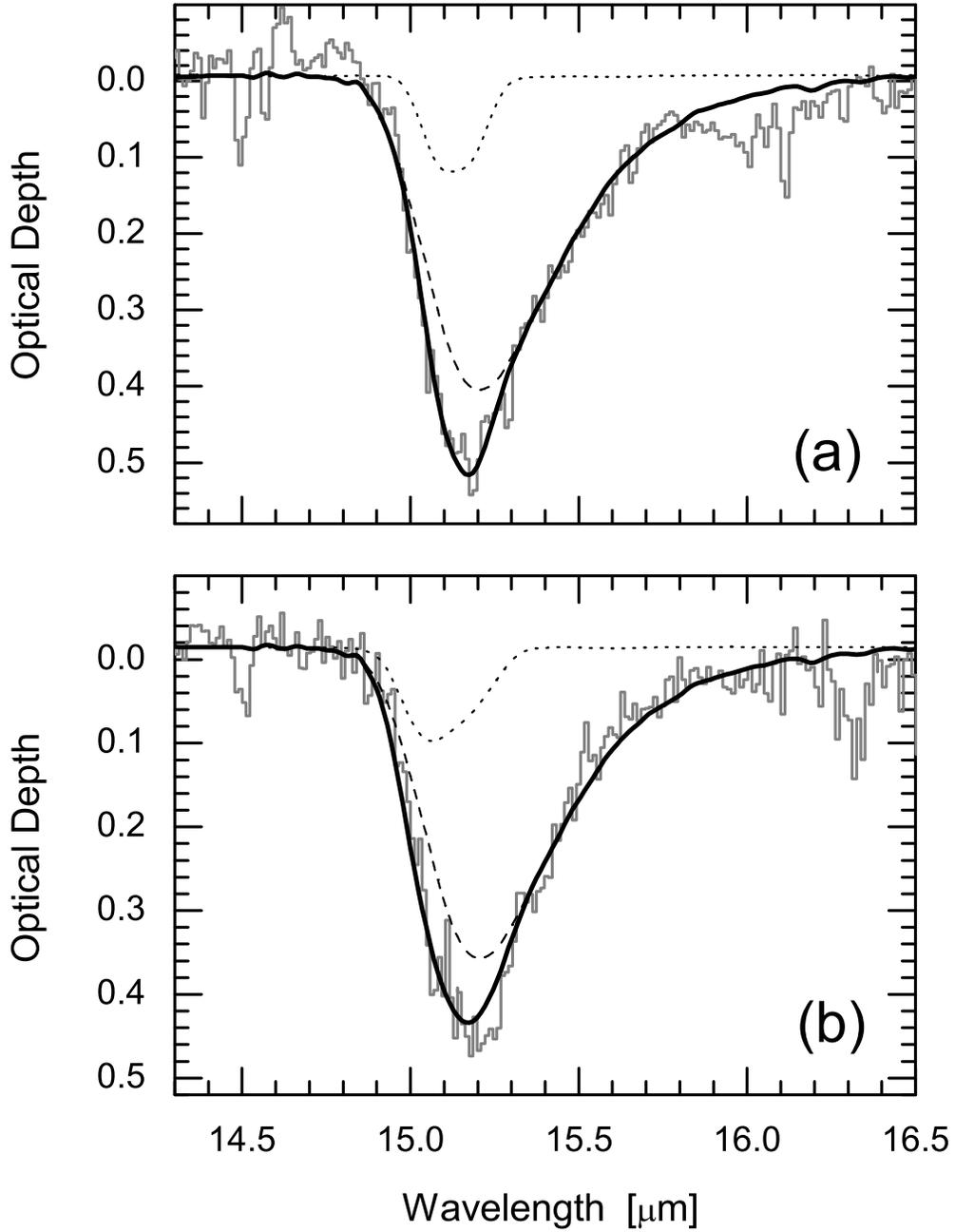}
\caption{Optical depth spectra of (a)~CK2 and (b)~Q21--1, with our best-fitting model to the \cod\ profile superposed in each case: polar component (dashed curves), nonpolar component (dotted curves), total (thick solid curves). The mixtures used in the fits are listed in Table~2.\label{fig2}}
\end{figure}

\clearpage

% Fig.3
\begin{figure}
\epsscale{0.9}
\plotone{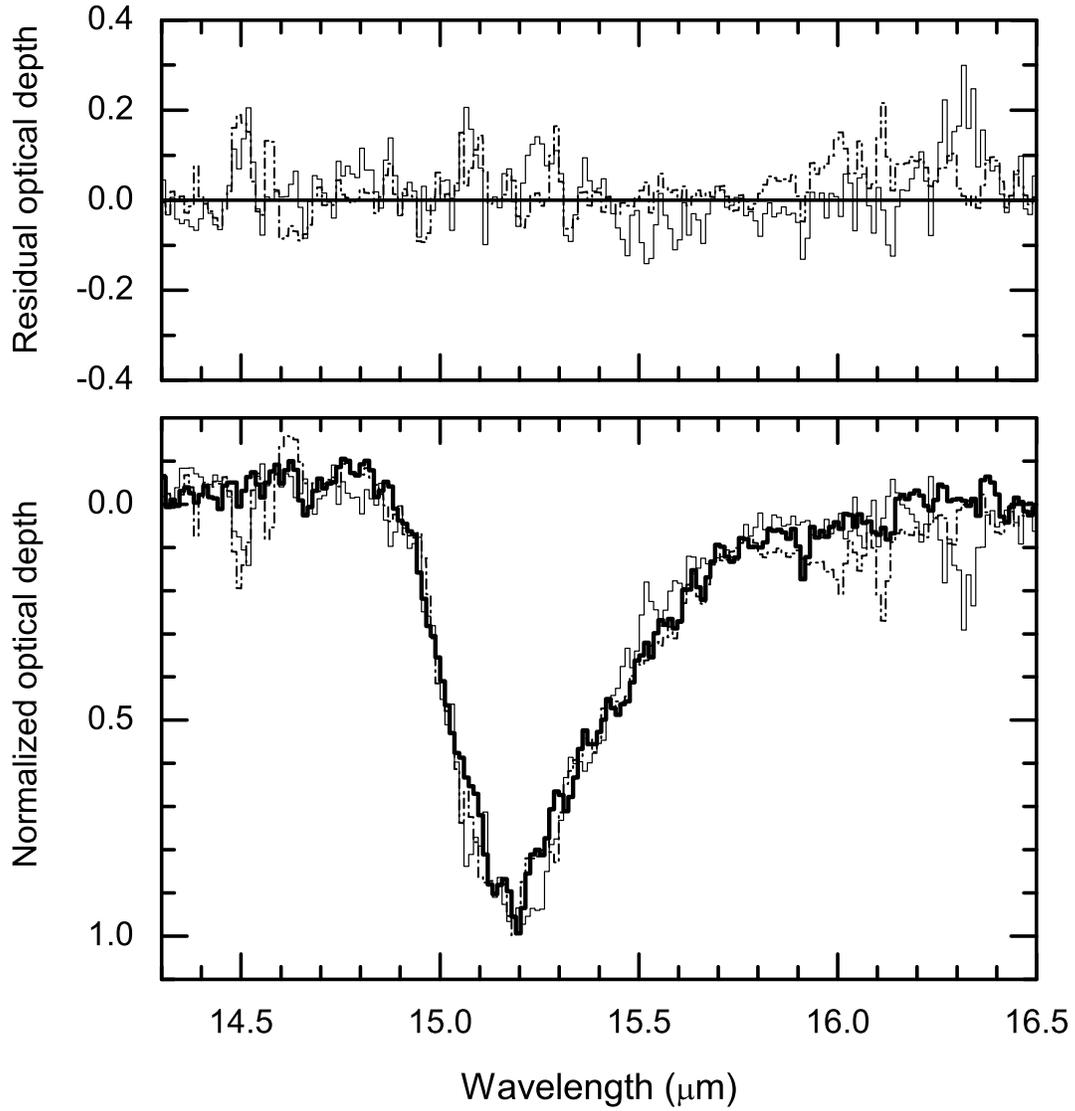}
\caption{Below: superposition of observed \cod\ profiles in all three field stars, normalized to unity at peak optical depth (Elias~16: thick solid curve; CK2: thin dotted curve; \hbox{Q21--1}: thin solid curve). Above: residuals obtained by subtracting the normalized profile of Elias~16 from those of CK2 and Q21--1 (dotted and solid curves, respectively). \label{fig3}}
\end{figure}

\clearpage

% Fig.4
\begin{figure}
\epsscale{0.85}
\plotone{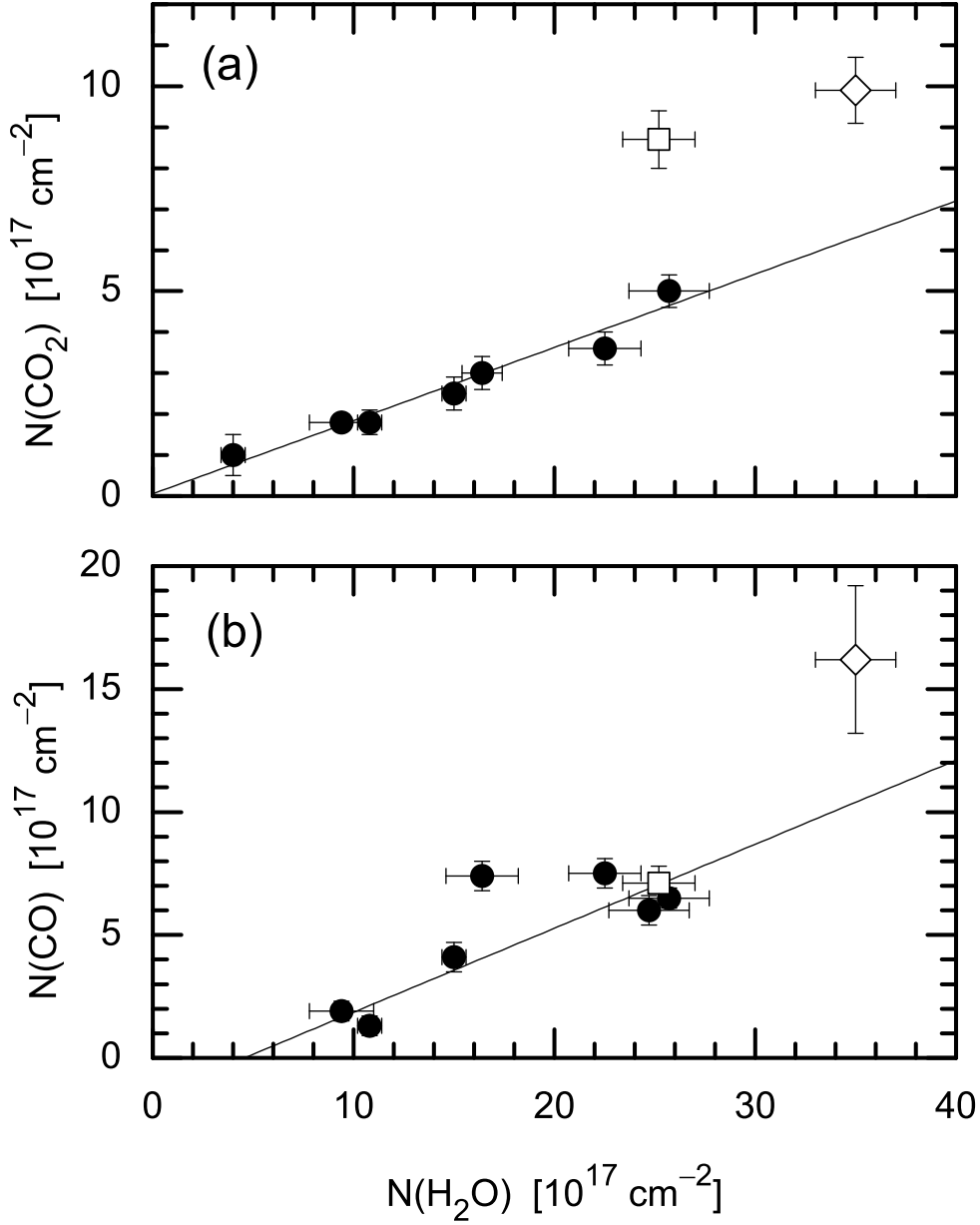}
\caption{Correlations of column densities for ices in field stars: (a)~$N$(\water) vs.\ $N$(\cod) and (b)~$N$(\water) vs.\ $N$(CO). Solid circles: Elias~16 and other Taurus stars from Whittet \etal\ (2007); open squares: Q21--1 (IC~5146); open diamonds: CK2 (Serpens). The diagonal line is the linear least-squares fit to Taurus stars only in each case. The slopes and intercepts (respectively) of the fits are (a)~$0.179\pm 0.042$, $0.055\pm 0.554$; (b)~$0.342\pm 0.026$, $-1.558\pm 0.450$. \label{fig4}}
\end{figure}

\clearpage

% Fig.5
\begin{figure}
\epsscale{0.85}
\plotone{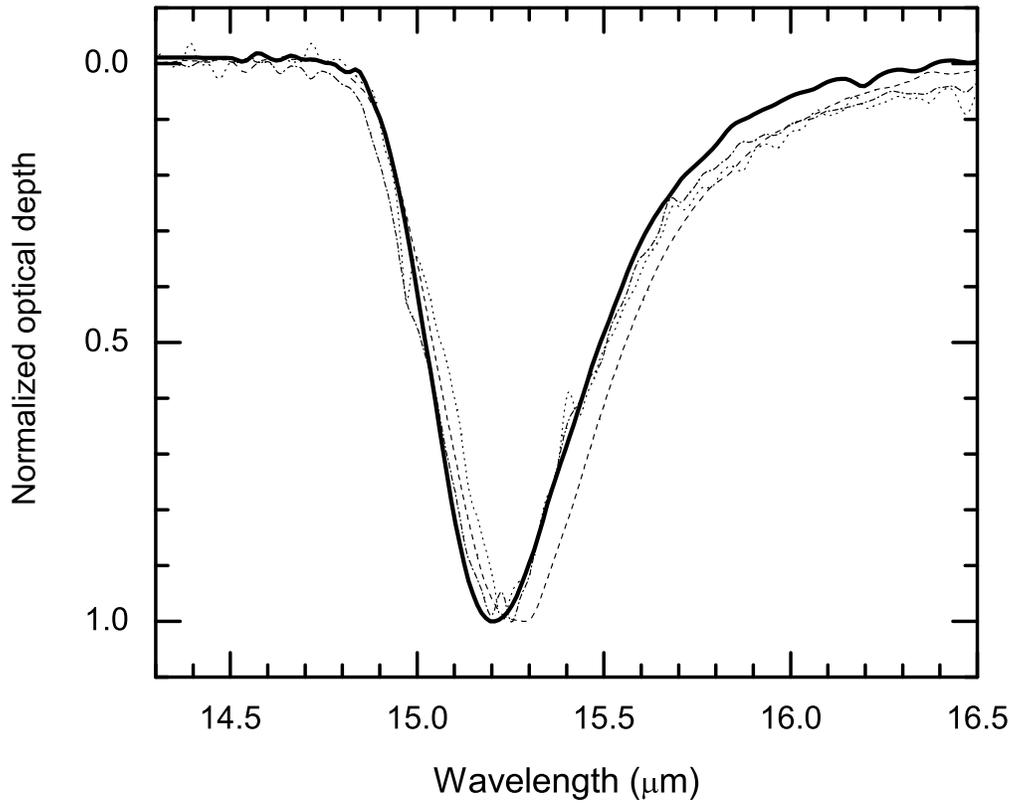}
\caption{Superposition of normalized \cod\ bending-mode profiles in laboratory ices at low temperature with differing \water:\cod\ ratios. The thick curve represents the polar mixture used in the fits shown in \fig 2 (\water:\cod:CO = 100:20:3 at 20~K); the other curves all represent \water:\cod\ mixtures at 10~K: 1:1 (dot-dash), 5:1 (dashed), and 100:14 (dotted). \label{fig5}}
\end{figure}

\clearpage

% Fig.6
\begin{figure}
\epsscale{0.85}
\plotone{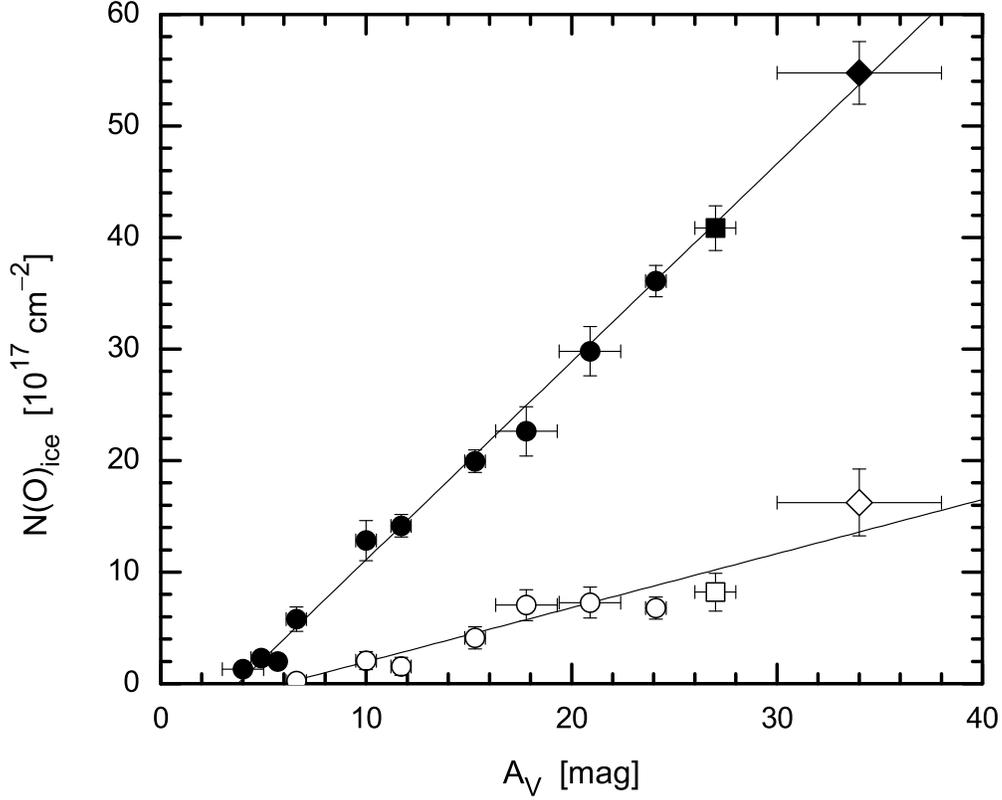}
\caption{Plots of the estimated column density $N$(O) of elemental oxygen in polar and nonpolar ice components (filled and open symbols, respectively) against the total visual extinction $A_V$. Taurus, Serpens and IC~5146 stars are represented by circles, diamonds and squares, respectively. The diagonal lines are linear least-squares fits to the polar and nonpolar data sets, combining results for all three clouds. The slopes and intercepts (respectively) of the fits are $0.175\pm 0.035$, $-6.646\pm 0.625$ (polar), and $-2.883\pm 0.764$, $0.485\pm 0.043$ (nonpolar). \label{fig6}}
\end{figure}

\clearpage

% Fig.7
\begin{figure}
\epsscale{0.85}
\plotone{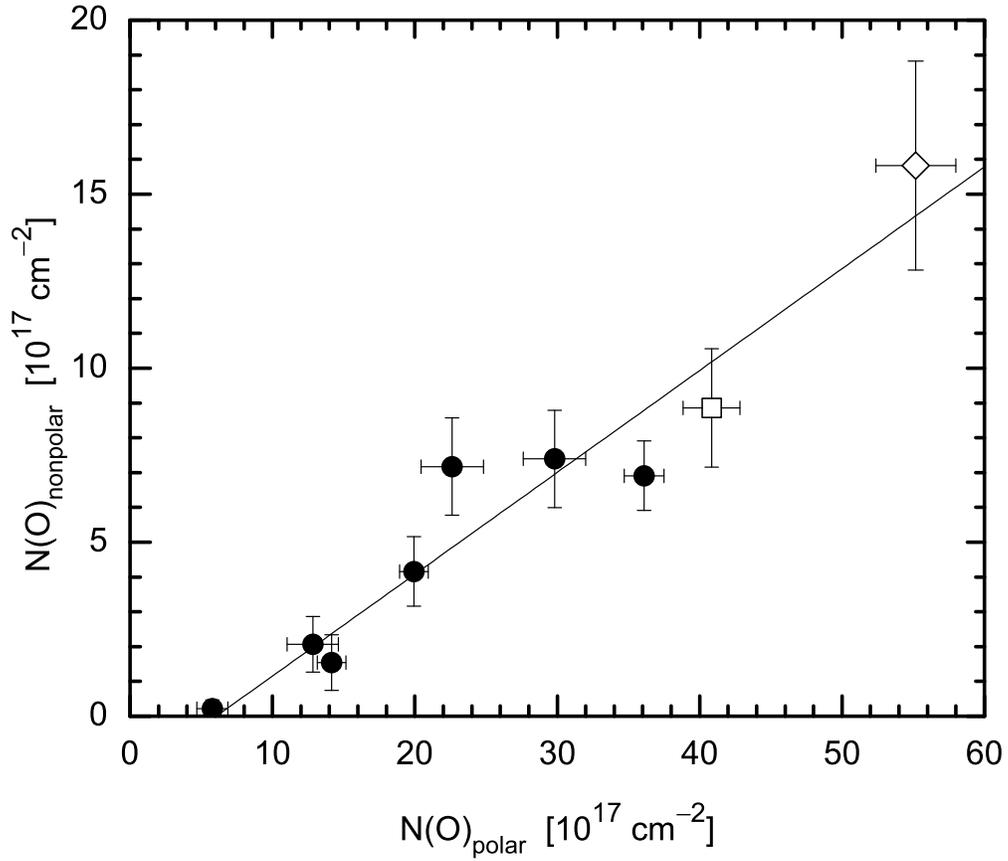}
\caption{Plot of the estimated column density $N$(O) of elemental oxygen in polar ice vs.\ that in nonpolar ice. Symbols have the same meaning as in \fig 4. The diagonal line is the linear least-squares fit to the combined data set for all three clouds. The slope and intercept of the fit have values $0.293\pm 0.036$ and $-1.774\pm 1.090$, respectively. \label{fig7}}
\end{figure}

\clearpage

\end{document}